\documentclass[
  reprint,
  aps,
  prb,
  superscriptaddress,
  amsmath,amssymb
]{revtex4-2}
\usepackage{xcolor}
\usepackage{bm}
\usepackage[squaren]{SIunits}
\usepackage{physics}
\usepackage{graphicx}
\usepackage[english]{babel}
\usepackage{orcidlink}

\usepackage{hyperref}
\hypersetup{colorlinks=true,linkcolor=blue,citecolor=blue,urlcolor=blue}
\begin{document}

\title{Experimentally controlling scattering of water waves in correlated disorder}

\author{Angélique Campaniello \orcidlink{0009-0006-5531-7031}}
\thanks{angelique.campaniello@espci.psl.eu} % -> *
\affiliation{Institut Langevin, ESPCI Paris, PSL University, CNRS, Paris, France}

\author{Rémi Carminati \orcidlink{0000-0002-5839-1820}}
\thanks{remi.carminati@espci.psl.eu} % -> †
\affiliation{Institut Langevin, ESPCI Paris, PSL University, CNRS, Paris, France}
\affiliation{Institut d’Optique Graduate School, Paris-Saclay University, 91127 Palaiseau, France}

\author{Marcel Filoche \orcidlink{0000-0001-8637-3016}}
\thanks{marcel.filoche@espci.psl.eu} % -> ‡
\affiliation{Institut Langevin, ESPCI Paris, PSL University, CNRS, Paris, France}

\author{Emmanuel Fort \orcidlink{orcid.org/0000-0003-2770-3753}}
\thanks{emmanuel.fort@espci.psl.eu} % -> ‡
\affiliation{Institut Langevin, ESPCI Paris, PSL University, CNRS, Paris, France}

\date{\today}

\begin{abstract}
Wave propagation in complex media is a universal problem spanning optics, acoustics, mechanics, and condensed matter physics. While disorder usually causes strong scattering, recent theory predicts that a special class of correlated disorder, known as stealthy hyperuniformity, can suppress scattering at long wavelengths, making a material transparent despite remaining structurally disordered and far from a simple homogenization regime. Experimental evidence of this remarkable transport regime within a medium has, however, remained limited. Here we report a direct, spatially resolved experimental observation of a transition between scattering and non-scattering wave transport induced by hyperuniform correlations. Using water waves as a model platform, we image both the amplitude and phase of the wavefield as it propagates through a two-dimensional disordered structure. This enables us to extract quantitative transport observables, including extinction lengths, statistical fluctuations, and energy-flow patterns, and to directly identify the boundary of the hyperuniform transparency regime. Our results provide a quantitative experimental validation of the transport regimes predicted for stealthy hyperuniform disorder and demonstrate that correlated disorder offers a powerful and practical route to control wave propagation in realistic systems across wave physics.

\end{abstract}

\maketitle

The propagation of waves in complex media is shaped by the interplay between disorder, interferences, and dimensionality~\cite{sheng_introduction_2006,sebbah_waves_2001,Akkermans_Montambaux_2007,carminati_principles_2021}. Controlling wave transport by engineering spatial correlations in the disorder has become an influential strategy across physics~\cite{Vynck23}, with implications for photonics~\cite{Wiersma13,Yu20,cao_harnessing_2022,Dawda25}, acoustics~\cite{derode_influence_2006,Gkantzounis17}, soft-matter~\cite{fraden_multiple_1990,rojas-ochoa_photonic_2004} and material science~\cite{garcia_photonic_2007}. Among many classes of correlated disorders, stealthy hyperuniform (SHU) structures play a prominent role. These systems suppress density fluctuations at large scale, causing their structure factor to vanish in a neighborhood of the origin in reciprocal space~\cite{Uche04,torquato_hyperuniform_2018}. This suppression eliminates scattering for sufficiently large wavelengths, yielding the remarkable property of transparency in materials that remain fully disordered~\cite{Leseur16}. More generally, SHU structures are expected to exhibit a full phase diagram of transport regimes, including transparency, bandgaps, Anderson localization, and enhanced absorption~\cite{florescu_designer_2009,Froufe_Perez17,Aubry20,Monsarrat22,sgrignuoli_subdiffusive_2022,Bigourdan19,Sheremet20}.

While theoretical understanding of SHU disorder has advanced rapidly, experimental validation has remained limited. Previous studies have probed global transmission properties, leaving the internal structure of the wavefield, and thus many of the distinctive predictions of correlated disorder, unresolved~\cite{haberko_transition_2020,Cheron22,Alhaitz23}. Qualitative intensity maps have been provided for electromagnetic waves in a cavity, showing evidence of the different transport regimes without quantitative metrics~\cite{Aubry20}. Spatially resolved measurements of wavefields, in both amplitude and phase,
have been achieved in microwave chaotic cavities~\cite{Kuhl_2005}, disordered
waveguides~\cite{Sebbah_2006}, diffusive systems in the optical
range~\cite{Bender_2020}, and hyperuniform photonic structures using
near-field imaging techniques~\cite{Granchi26}. In this work, we use water waves to probe wave transport in two-dimensional SHU disorder, leveraging the ability of this platform to provide direct, non-invasive, spatially resolved measurements of the complex wavefield over an extended region including the inside of the disordered medium and over a broad wavelength range~\cite{Ohl01,Hu05,apffel_experimental_2022}. From these maps we extract quantitative observables that diagnose the different transport regimes, enabling a direct comparison with theoretical expectations for SHU structures. Our work also demonstrates, for the first time, the impact of SHU disorder on mechanical surface waves~\cite{Zhu24}. A salient feature of these waves is that water constitutes an absorbing background medium, which introduces dissipation that competes with correlation-induced transparency. This setting allows us to examine not only the ideal predictions of SHU disorder but also their robustness in lossy environments.

To identify the transition between suppressed and strong scattering regimes, we employ complementary observables, such as the extinction length, which quantifies the exponential attenuation of the average field~\cite{carminati_principles_2021}, or current-density contours inside the medium, which map the internal flow of energy and visualize how SHU correlations shape transport pathways. By combining these observables with precise control over the spatial configurations of scatterers, we experimentally identify the boundaries of the stealth regime, demonstrate the suppression of scattering, and quantify how dissipation modifies the transparency window. These results constitute the first direct in-medium experimental characterization of wave transport in SHU disorder in an absorbing background. More broadly, they show how structural correlations in disordered media can be exploited to control wave propagation even in realistic, non-ideal environments, expanding the design principles for disorder-based metamaterials.

\section*{Scattering and absorption in hyperuniform disorder}

We consider gravity-capillary waves on the surface of a water bath to study wave scattering in two dimensions. For an inviscid and incompressible fluid, and for small vertical displacements, monochromatic waves at an angular frequency~$\omega$ can be characterized by a scalar amplitude $\psi(\vb{r},t) = \mathrm{Re} [\phi(\vb{r})\exp(-i\omega t)]$, with $\phi(\vb{r})$ satisfying the Helmholtz equation. For plane waves $\phi(\vb{r})=A \exp(i\vb{k}\cdot \vb{r})$, the dispersion relation takes the form~\cite{Landau}
\begin{equation}
\omega^2 = \left (gk + \frac{\gamma}{\rho_w} k^3 \right ) \tanh(kh) \, ,
\label{eq:dispersion}
\end{equation}
where $\rho_w$ is the water density, $\gamma$ the water surface tension coefficient, $g$ the acceleration of gravity, $h$ the liquid bath depth and $k=|{\bf k}|$.

Upon interaction with a disordered medium, waves undergo multiple scattering, and a key parameter characterizing their transport is the scattering mean free path $\ell_s$. For an uncorrelated set of discrete scatterers, the mean free path is simply $\ell_s=(\rho\sigma_s)^{-1}$, with $\rho$ the number density of scatterers and $\sigma_s$ the scattering cross-section of an individual scatterer. In the presence of statistical correlations in the scatterers' positions, scattering is influenced by interferences encoded in the structure factor~\cite{Vynck23,carminati_principles_2021}
\begin{equation}
S(q) = \frac{1}{N} \left \langle \left | \sum_{j=1}^N \exp(-i{\bf q}\cdot {\bf r}_j) \right |^2 \right \rangle \, ,
\label{eq:S_factor}
\end{equation}
where ${\bf r}_j$ is the position of scatterer number $j$ in a set of $N$ scatterers, and the brackets denote an ensemble average over realizations of disorder. We assume statistical isotropy so that the structure factor depends only on $q=|{\bf q}|$. In a correlated set in two dimensions the scattering mean free path becomes
\begin{equation}
\frac{1}{\ell_s} =\rho \int_{0}^{2\pi} S(q) \, |f(q)|^2 d\theta \,,
\label{eq:MFP_S}
\end{equation}
where $q=2k_{\mathrm{eff}}|\sin(\theta/2)|$, with $k_{\mathrm{eff}}$ the real part of the effective wavenumber characterizing the propagation of the average field in the multiple scattering regime~\cite{Vynck23,carminati_principles_2021}. When $\ell_s$ remains large compared to the wavelength $\lambda=2\pi/k$, we can assume $k_{\mathrm{eff}}\simeq k$, with $k$ the wavenumber in the background medium solution to \eqref{eq:dispersion}. In this expression $f(q)$ is the scattering amplitude of an individual scatterer such that $\int |f(q)|^2 d\theta = \sigma_s$, and $S(q)$ is the structure factor with the forward scattering contribution subtracted~\cite{carminati_principles_2021}.

Stealthy hyperuniform distributions of scatterers are characterized by a structure factor satisfying $S(q) = 0$ for $q \leq K$. In this case, we find from Eq.~(\ref{eq:MFP_S}) that $\ell_s \to \infty$ for $k \leq K/2$. Physically, the constraint on the structure factor induces spatial correlations in the positions of the scatterers that suppress density fluctuations at large scales, thus suppressing scattering at large wavelengths or equivalently small wavevectors. Therefore, by increasing the wavenumber $k$ across a spectral region including the threshold wavenumber $K/2$, the SHU medium is expected to exhibit a transition from non-scattering to strongly scattering.

\begin{figure}[h]
\centering
\includegraphics[width=1.0\linewidth]{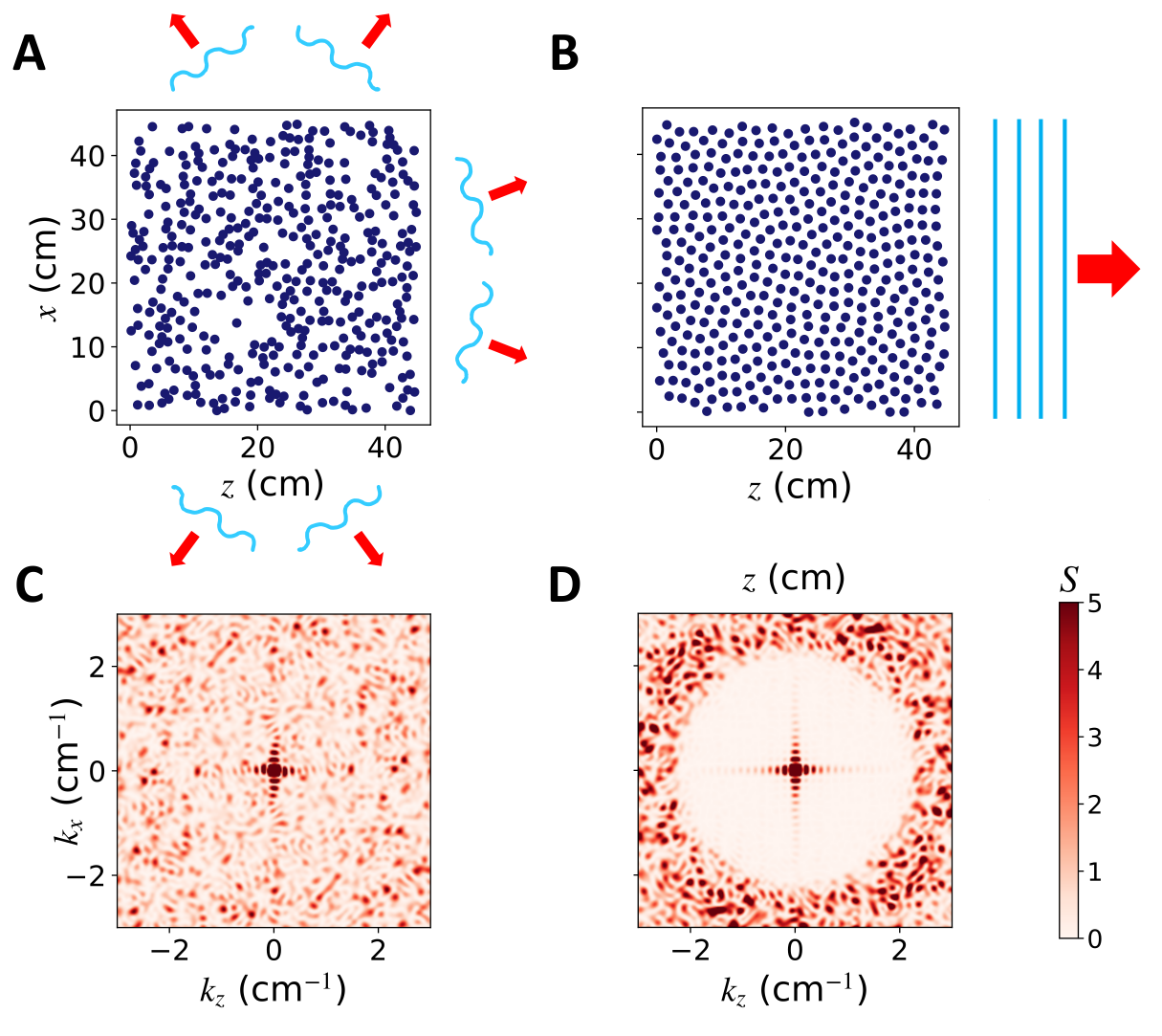}
\caption{Disordered patterns used in the experiments. Uncorrelated pattern (A) and SHU pattern (B), together with their corresponding structure factors $S$ (C) and (D). The SHU region around $q=0$ is clearly visible in (D). The uncorrelated medium scatters the incoming wave in all directions (red arrows in frame A), whereas the SHU medium allows waves below a threshold frequency to propagate without scattering, (frame B). Note that the number density of scatterers is the same in both cases.}
\label{fig:structure factor}
\end{figure}

The degree of spatial correlations in SHU disorder scales with the size $K$ of the constrained region, and is usually measured by the parameter $\chi$ such that $K = \sqrt{16\pi \rho \chi}$, with $\chi=0$ corresponding to an uncorrelated disorder and $\chi = 1$ to a periodic crystal~\cite{Uche04,torquato_hyperuniform_2018}. Here we chose $\chi=0.5$, which corresponds to highly correlated albeit disordered configurations. The value of $K$, and therefore the spectral range of the non-scattering region, is adjusted using the density $\rho$ as a parameter. Examples of fully disordered and SHU patterns are shown in Fig.~\ref{fig:structure factor}A and B, together with their corresponding structure factors in Fig.~\ref{fig:structure factor}C and D. The SHU patterns were generated following the optimization procedure described in Ref.~\cite{Monsarrat22} (see Materials and Methods and SI Appendix, Fig. S2). We note that SHU patterns do not require ensemble averaging to reveal their small-$q$ suppression: Each single realization individually satisfies the constraint $S(q)=0$ for $q \le K$ (see Fig.~\ref{fig:structure factor}D). By contrast, uncorrelated disorder exhibits configuration to configuration fluctuations of $S(q)$ over the entire $q$ plane (Fig.~\ref{fig:structure factor}C).

For water waves, attenuation due to viscosity cannot be neglected. Exponential attenuation of the wave amplitude is described by a complex-valued wavenumber $k^\star=k+i\beta$, where $\beta = 4 \eta k^2/(\rho_w v_g)$, with $\eta$ the dynamical viscosity and $v_g$ the group velocity. In practice, the observed decay rate~$\beta$ exceeds the theoretical bulk prediction, due to additional sources of dissipation, such as surface contamination and friction at the bottom and lateral walls of the water tank. In principle, dissipation in the background should result in a correction factor in the summation in \eqref{eq:S_factor}, creating a cutoff $\ell_a=1/\beta$ in the distance between pairs of scatterers capable of producing interferences. In our experiment, this cutoff does not significantly hinder the ability to observe the crucial role played by the hyperuniform disorder, since in the frequency range \unit{3.5-10}{Hz} that is used, the condition $\ell_a \gg \lambda$ is always satisfied. Indeed, $\ell_a$ ranges from \unit{19}{cm} to \unit{147}{cm}, while the wavelength $\lambda$ ranges from \unit{2.3}{cm} to \unit{14}{cm}. Even in the presence of viscous dissipation, we can safely conclude that changes in the scattering mean free path $\ell_s$ due to hyperuniform correlations are well described by Eqs.~(\ref{eq:S_factor}) and~(\ref{eq:MFP_S}).

\section*{Experimental setup}

\begin{figure}[h!]
    \centering
    \includegraphics[width=\linewidth]{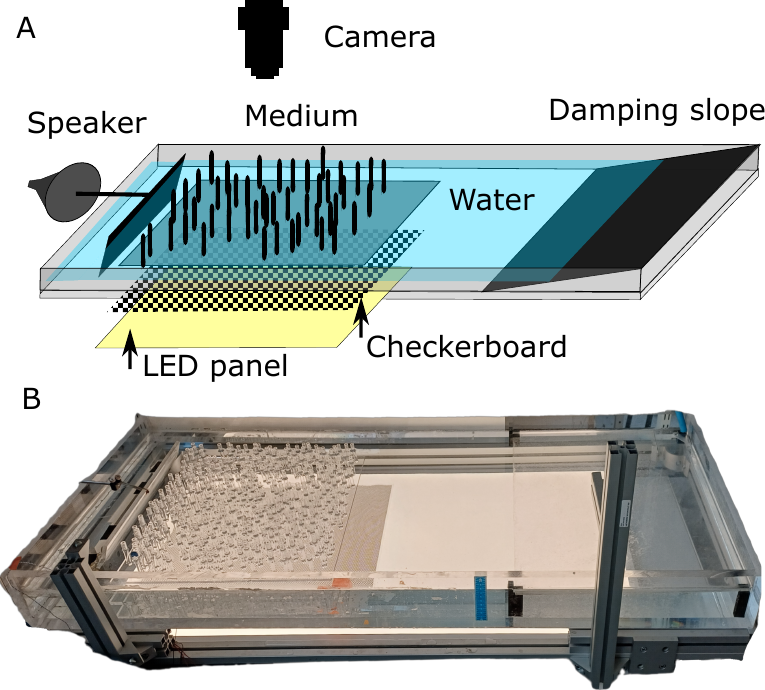}

    \caption{(A) Schematic representation of the experiment. A loudspeaker drives a 40-cm paddle to generate monochromatic surface waves in a 150 × 60 cm water tank. A camera placed above the tank records the deformation of a checkerboard located beneath the transparent bottom, enabling reconstruction of the water surface elevation (wavefield). A sloped absorbing beach suppresses unwanted reflections. (B) Photograph of the actual setup, showing the tank, the paddle, the illumination system, and a hyperuniform medium.}
    \label{fig:setup}
\end{figure}

The experimental setup consists of a 150 × 60 cm water tank filled to a depth of 7~cm, as shown in Fig.~\ref{fig:setup}. Surface waves are generated by a 40-cm-wide paddle driven by a loudspeaker, producing monochromatic waves with frequencies between 3.5 and 10~Hz, which allows us to probe both regimes with suppressed and strong scattering in SHU media. Over this range, the wavelength varies from 2.3 to 14~cm, and the paddle is sufficiently wide to generate quasi-plane waves in the area of observation (SI appendix, Movie~S1). 

The wavefield is measured using a camera positioned 2~m above the tank, and imaging a checkerboard pattern placed beneath the transparent bottom. 
The deformation of the checkerboard is converted into a quantitative map of the surface elevation through Fast Fourier checkerboard demodulation, which provides access to both the amplitude and phase of the wavefield throughout the observation region, with approximately 20~µm vertical and 0.3~mm lateral resolution~\cite{Wildeman18}. After spatio-temporal filtering at the excitation frequency and masking of the scatterers, the complex amplitude $\phi({\bf r})$ is extracted using a lock-in procedure. 

Disordered media are produced by cutting circular holes with lasers in a transparent 45 × 45~cm and 8~mm-thick PMMA plate in which vertical cylindrical PMMA scatterers (diameter 1 cm) are inserted (see Materials and Method). The scatterers are non-resonant and subwavelength across the full frequency range. Two media of identical density $\rho$ = \unit{0.2}{cm^{-2}}, each consisting of $N$=401 scatterers, are produced using i) an uncorrelated distribution obtained by random sequential addition, and ii) a SHU distribution generated with $\chi=0.5$, corresponding to $K \simeq 2.24$~cm$^{-1}$. The transparency threshold in the SHU medium is predicted to occur at $k = K/2 \simeq 1.12$~cm$^{-1}$, or a frequency of~\unit{6}{Hz}.
Each plate can rotate by 90°, providing four different realizations of disorder in both cases. 

Viscous attenuation is characterized separately by measuring the absorption length $\ell_a$  of freely propagating plane waves in the absence of scatterers. This background absorption can be subtracted from the extinction measurements to isolate the effects of the structural disorder.

\section*{Scattering cross section of a single scatterer}

\begin{figure}[h!]
\centering
\includegraphics[width=\linewidth]{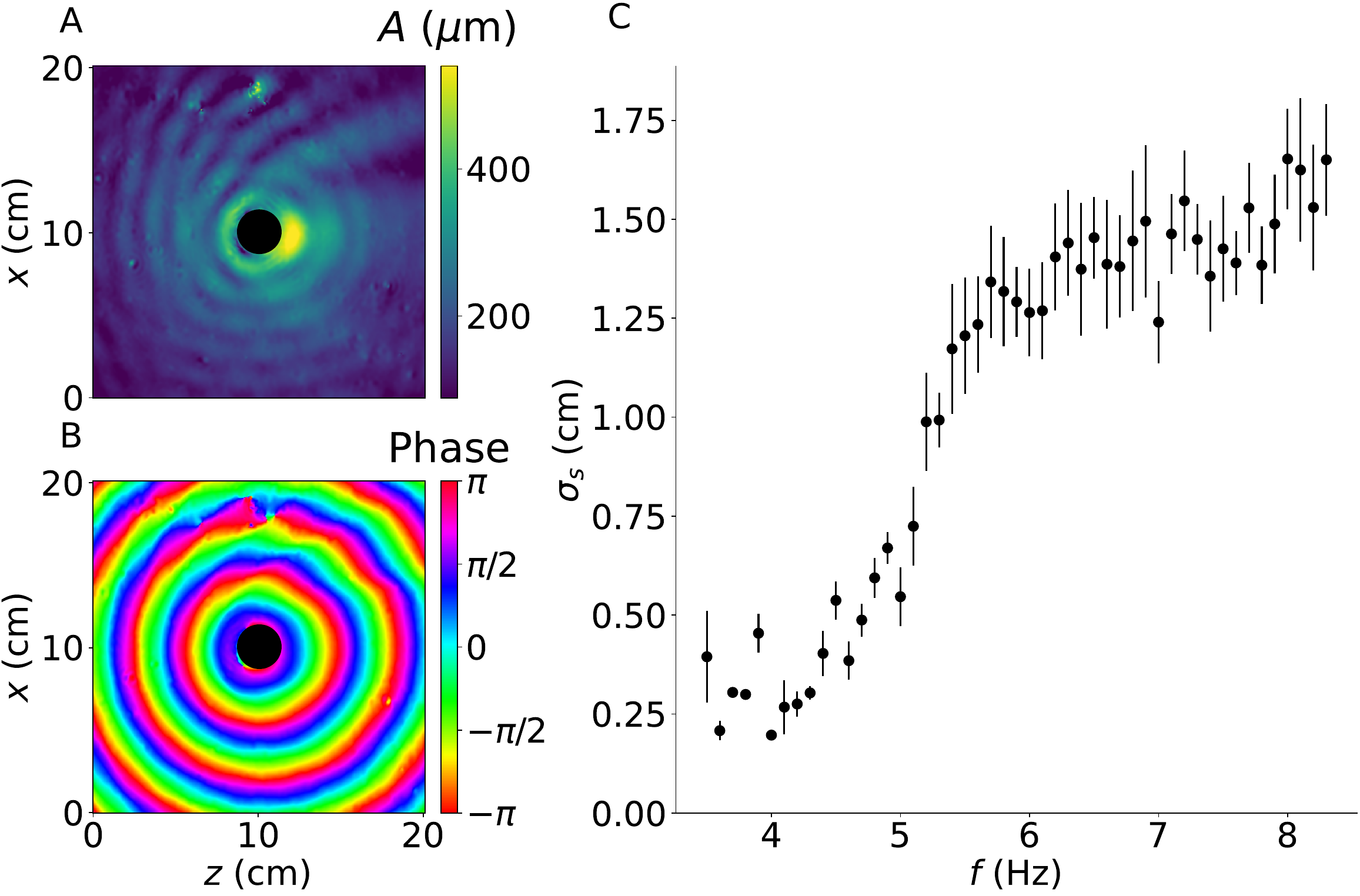}
\caption{Analysis of a single scatterer. (A)~Amplitude of the extracted scattered field at an excitation frequency of~\unit{7.5}{Hz}. (B)~Phase of the scattered field at~\unit{7.5}{Hz}. (C)~Measured scattering cross-section~$\sigma_s$, averaged over several values of the radius [see~\eqref{eq:sigma_s}] versus the excitation frequency~$f=\omega/2\pi$ for a single scatterer.}
\label{fig:sigma_s}
\end{figure}

We characterize the scattering cross-section of a single scatterer consisting of a vertical cylinder with radius $a$ subjected to an incident monochromatic plane wave $\phi_i(\vb{r})$. The complex amplitude of the total field is decomposed into incident and scattered contributions, $\phi(\vb{r})=\phi_i(\vb{r})+\phi_s(\vb{r})$. Before extracting the amplitude and phase from the measurement, we perform a spatio-temporal filtering (see Materials and Methods section). The scattered amplitude~$\abs{\phi_s}$ and phase~$\arg(\phi_s)$, are displayed in Fig.~\ref{fig:sigma_s}A and B. The phase maps reveal nearly perfect circular contours, while the amplitude decreases radially from the cylinder. These features are consistent with a predominantly monopolar scattering response. The scattering cross-section is extracted from the scattered amplitude using the relation
\begin{equation}\label{eq:sigma_s}
    \sigma_s = \exp(2R/\ell_a) \; \frac{\int_0^{2\pi} \abs{\phi_s(\theta)}^2 R\,{\rm d}\theta}{\abs{\phi_i}^2} \,,
\end{equation}
where the integral is performed along a circle of radius~$R$ centered on the scatterer. The denominator normalizes the scattered intensity by the local incident intensity. The prefactor~$\exp(2R/\ell_a)$ corrects for viscous attenuation in the background, with the attenuation length $\ell_a$ obtained experimentally. Another approach could have involved measuring the extinction cross-section based on the forward scattering amplitude using the optical theorem. In an absorbing medium, the method described here, which uses angular integration, proved to be more robust, with the advantage of directly yielding the scattering cross section. Theoretically, one would expect a cross-section independent of the radius $R$, provided that the scattered field has reached its asymptotic form. In the far field, $\phi_s(\mathbf{r}){\sim}f(\theta) \exp(ikr)\exp(-r/\ell_a)/\sqrt{r}$, and the cross-section is found to be independent of distance. In practice, this can be checked by computing $\sigma_s$ for different values of $R$. Figure~\ref{fig:sigma_s}C displays the average value of $\sigma_s$ together with its standard deviation (error bars) versus the excitation frequency. The main contribution to this standard deviation comes from the residual incident field remaining after the extraction procedure. The standard deviation increases at high frequency since, in this regime, the ratio between the scattered and incident fields becomes smaller due to viscous attenuation. Over the explored frequency range, we observe a smooth increase of the scattering cross-section~$\sigma_s$ with no resonant peaks, even though $ka$ ranges from 1 to 3, meaning that the cylinders are in the Mie scattering regime.

\section*{Evidence of suppression of multiple scattering in a stealthy hyperuniform medium}

We now study the propagation of surface water waves in a medium made of a disordered set of identical scatterers. We quantitatively characterize the transport regime in both uncorrelated and SHU configurations.

\subsection{Extinction length}

\begin{figure}[h!]
\centering
\includegraphics[width=\linewidth]{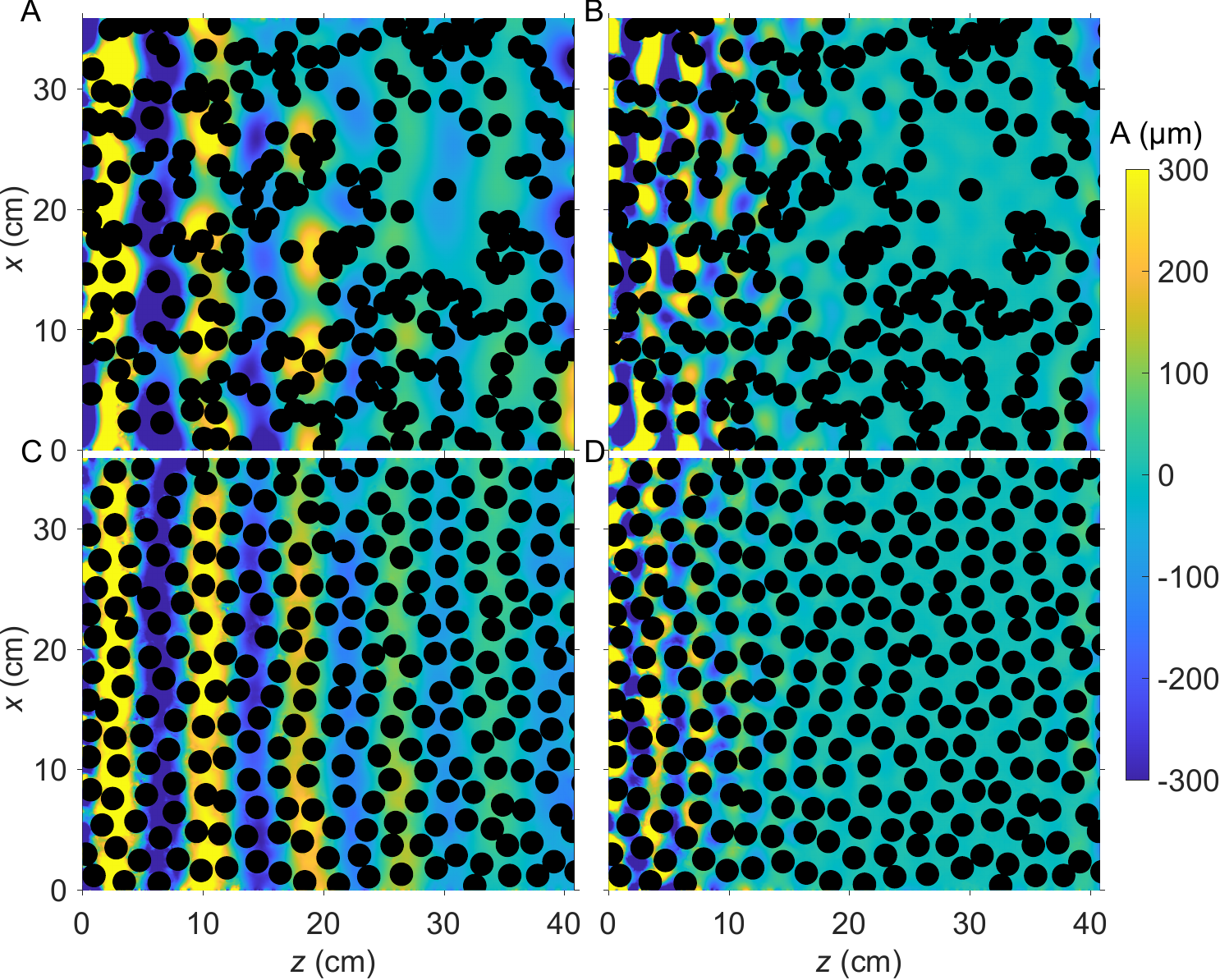}
\caption{Maps of measured wavefields. (A)~Uncorrelated medium excited at frequency~$f=\unit{5}{Hz}$ by a plane wave propagating from left to right along the $z$ direction. (B)~Uncorrelated medium at $f=\unit{8}{Hz}$. (C)~SHU medium at frequency $f=\unit{5}{Hz}$. The unperturbed wavefield reveals the absence of scattering. (D)~SHU medium at $f=\unit{8}{Hz}$ above the critical value. Colors indicate the water surface elevation, and the black dots correspond to the positions of the scatterers.
}
\label{fig:wavefield}
\end{figure}

\begin{figure}[h!]
\centering
\includegraphics[width=\linewidth]{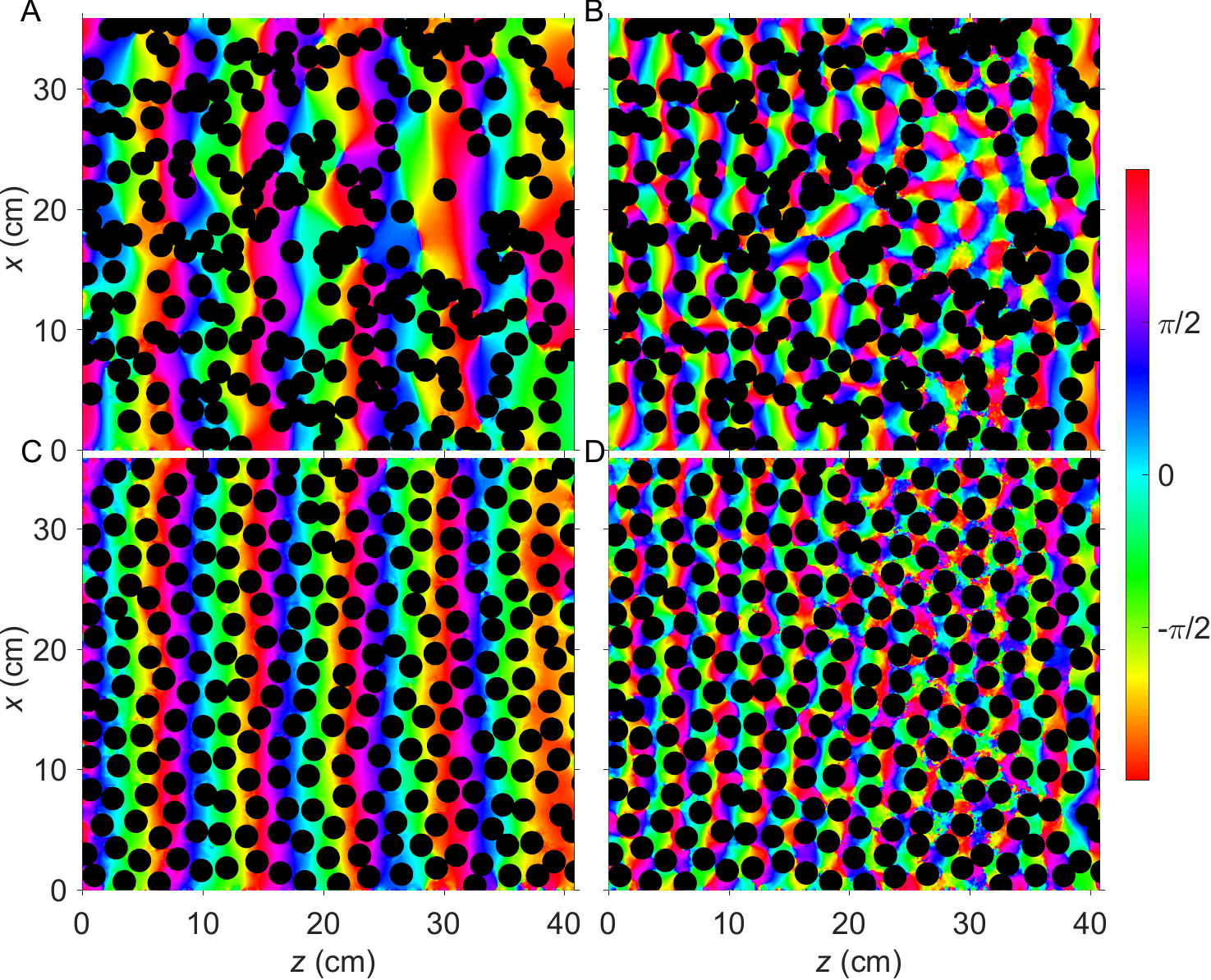}
\caption{Maps of the phase of the measured wavefields. 
(A)~Phase map of an uncorrelated medium excited by a plane wave propagating from left to right along the $z$ direction at frequency $f=\unit{5}{Hz}$. 
(B)~Uncorrelated medium at $f=\unit{8}{Hz}$. 
(C)~SHU medium at $f=\unit{5}{Hz}$, showing an essentially unperturbed wavefront. 
(D)~SHU medium at $f=\unit{8}{Hz}$, above the critical frequency, where scattering becomes significant. 
Colors represent the phase of the wavefield, and black dots indicate the positions of the scatterers.}

\label{fig:phase map}
\end{figure}

Figure~\ref{fig:wavefield} displays the wavefields measured inside the disordered medium at two different excitation frequencies, \unit{5}{Hz} and \unit{8}{Hz} (SI Appendix, Movie~S2). The incident field is a plane wave propagating from left to right along the $z$~direction. For the uncorrelated medium (Fig.~\ref{fig:wavefield}A and \ref{fig:wavefield}B), we observe a deformed wavefront at both frequencies, as a consequence of scattering, in agreement with the phase distortions observed in the corresponding phase maps (Fig.~\ref{fig:phase map}A and \ref{fig:phase map}B). For the SHU medium, the wave front remains unchanged at~\unit{5}{Hz} (Fig.~\ref{fig:wavefield}C), revealing the absence of scattering, further confirmed by the preservation of planar phase fronts in the corresponding phase map (Fig.~\ref{fig:phase map}C). At this frequency the wavenumber $k=\unit{0.73}{cm^{-1}}$ is smaller than~$K/2$, corresponding to the non-scattering region. At~\unit{8}{Hz}, however (Fig.~\ref{fig:wavefield}D), the wavenumber is $k=\unit{1.77} {cm^{-1}} > K/2$, and the wave is scattered even in the SHU medium. This is also evidenced by the phase distortions (Fig.~\ref{fig:phase map}D). Changes in scattering strength versus frequency are also observed on amplitude and phase maps over the full spectral range (SI Appendix, Fig.~S1) and on spatial Fourier spectra (SI Appendix, Movie~S3 and Fig.~S3).

To quantitatively characterize the change in the scattering strength, a relevant parameter is the extinction length $\ell_e$, defined as the decay length of the average field $\langle \phi \rangle \sim \exp(-z/2\ell_e)$~\cite{sheng_introduction_2006,Akkermans_Montambaux_2007,carminati_principles_2021}. In practice, the averaging process involves four configurations of disorder obtained by rotating the square PMMA plate holding the scatterers by 90 degrees, and a spatial averaging of the field amplitude along the transverse direction (the $x$-axis in Fig.~\ref{fig:wavefield}). From the averaged field, a fit of the exponential decay allows us to deduce $\ell_e$. We subtract the effect of viscous attenuation in the background medium by independently measuring the absorption length $\ell_a$ in the absence of scatterers, and use the coefficient $\alpha_e = 1/\ell_e-1/\ell_a$ as a measure of the extinction strength due to structural disorder. Note that this coefficient may include a residual dissipative process induced by friction at the scatterers boundaries, which is supposed to be weakly affected by spatial correlations. In the following, we use this coefficient to probe changes in the scattering strength induced by spatial correlations, and refer to it as effective scattering coefficient.

\begin{figure}[h]
\centering
\includegraphics[width=0.8\linewidth]{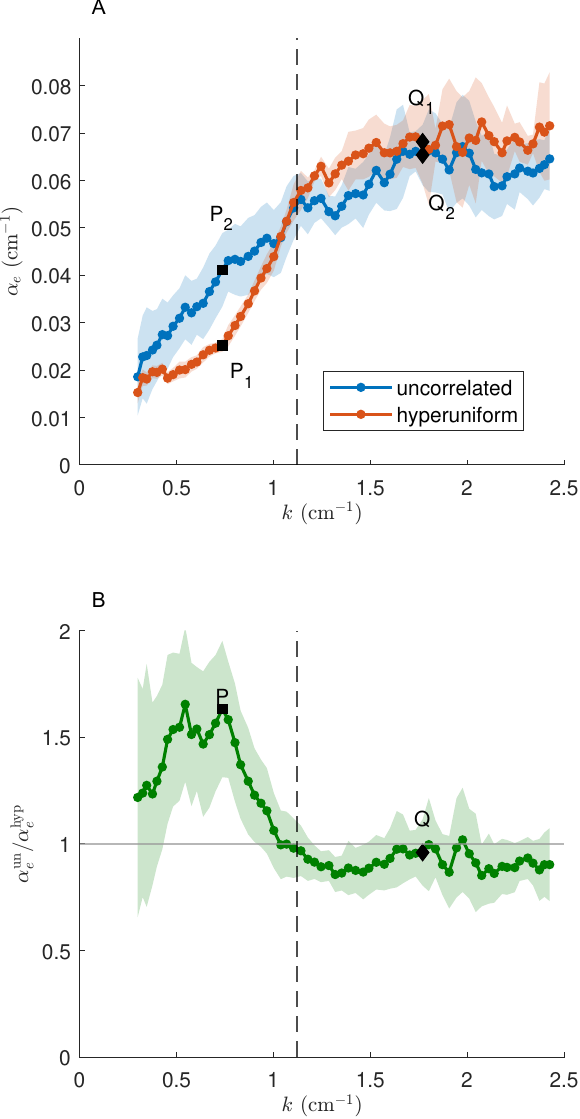}
\caption{Effective scattering coefficient. (A)~Measured effective scattering coefficient $\alpha_e$ versus the wavenumber $k$ for a SHU medium (blue) and an uncorrelated medium (red) with density $\rho= \unit{0.2}{cm^{-2}}$. The theoretical threshold wavenumber $k_{\rm hyp} = K/2 = \unit{1.12}{cm^{-1}}$ is indicated by the vertical dashed line. (B)~Ratio of the effective scattering coefficients $\alpha_e^\text{un}/\alpha_e^\text{\rm hyp}$ versus the wavenumber $k$ showing a clear transition at the threshold wavenumber. The markers $P_1$ and $P_2$ correspond to the hyperuniform and uncorrelated configurations at $f=\unit{5}{Hz}$, respectively, while $Q_1$ and $Q_2$ are at $f=\unit{8}{Hz}$, respectively. The associated wave fields are shown in Fig.~\ref{fig:wavefield}.}
\label{fig:l_e}
\end{figure}

Figure~\ref{fig:l_e}A displays the effective scattering coefficient $\alpha_e$ versus the incident wavenumber~$k$ for the uncorrelated (blue) and the SHU media (red). The error bars, represented by the blue and red shaded regions, correspond to the statistical root mean square resulting from the two-steps averaging process described above. The critical value of the wavevector $K/2=\unit{1.12}{cm^{-1}}$ separating the non-scattering and the scattering regions for the SHU medium is indicated by the vertical black dashed line. We observe a clear transition: for $k\leq K/2$, extinction in the SHU medium is significantly lower than in the uncorrelated medium, whereas for $k> K/2$, extinction increases substantially and is even larger in the SHU medium. The transition is even more visible in Fig.~\ref{fig:l_e}B where we plot the ratio $\alpha_e^\text{un}/\alpha_e^\text{\rm hyp}$ versus the wavenumber, with $\alpha_e^\text{un}$ and $\alpha_e^\text{\rm hyp}$ the effective scattering coefficients for the uncorrelated and SHU media, respectively. Below the threshold wavenumber~$K/2$, the ratio is of the order of 1.5, indicating the strong reduction of scattering in the SHU medium, while the ratio drops to $\simeq 0.9$ for $k > K/2$.

\subsection{Fluctuations of the scattering strength}

\begin{figure}[h]
\centering
\includegraphics[width=0.8\linewidth]{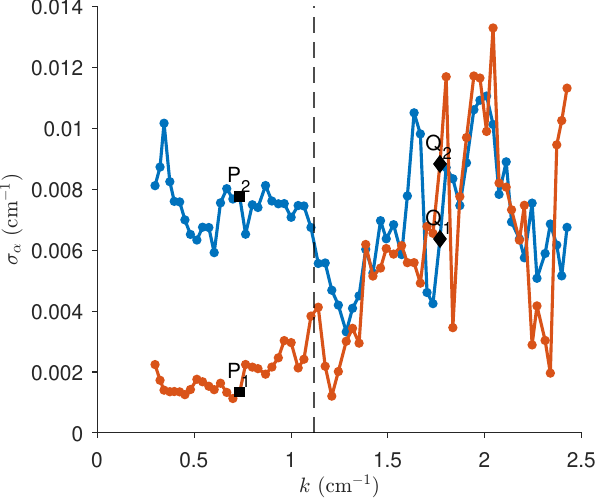}
\caption{Standard deviation of the extinction lengths. Standard deviation of the extinction lengths versus the wavenumber $k$ for a SHU medium (blue) and an uncorrelated medium (red). The theoretical critical wavenumber $k_{\rm hyp} = K/2 = \unit{1.12}{cm^{-1}}$ is indicated by the vertical black dashed line.}
\label{fig:l_e std}
\end{figure}

The error bars in Fig.~\ref{fig:l_e}A show a substantial reduction of fluctuations for the SHU medium in the non-scattering region $k \leq K/2$. This is clearly seen in Fig.~\ref{fig:l_e std}, which displays the standard deviation~$\sigma_\alpha$ of the effective scattering coefficient as a function of wavenumber. The effective scattering coefficient of the uncorrelated medium (blue curve) exhibits a large standard deviation for all values of the wavenumber. In contrast, for the SHU medium, the standard deviation nearly vanishes in the non-scattering region. 

The absence of sample-to-sample fluctuations in the non-scattering regime for the SHU medium stems from the fact that each realization individually satisfies $S(q)=0$ for $q \le K$ (the dark blue region in Fig.~\ref{fig:structure factor}D is independent of the statistical realization of disorder). Outside this region, or for an uncorrelated pattern, the random structure of $S(q)$ is responsible for the fluctuations of the effective scattering coefficient. As a result, with respect to scattering suppression, a single SHU sample faithfully reflects the behavior of the ensemble, making $\alpha_e^\text{hyp}$ a self-averaging quantity in the non-scattering region.

\subsection{Flux lines}

\begin{figure}[h!]
\centering
\includegraphics[width=1\linewidth]{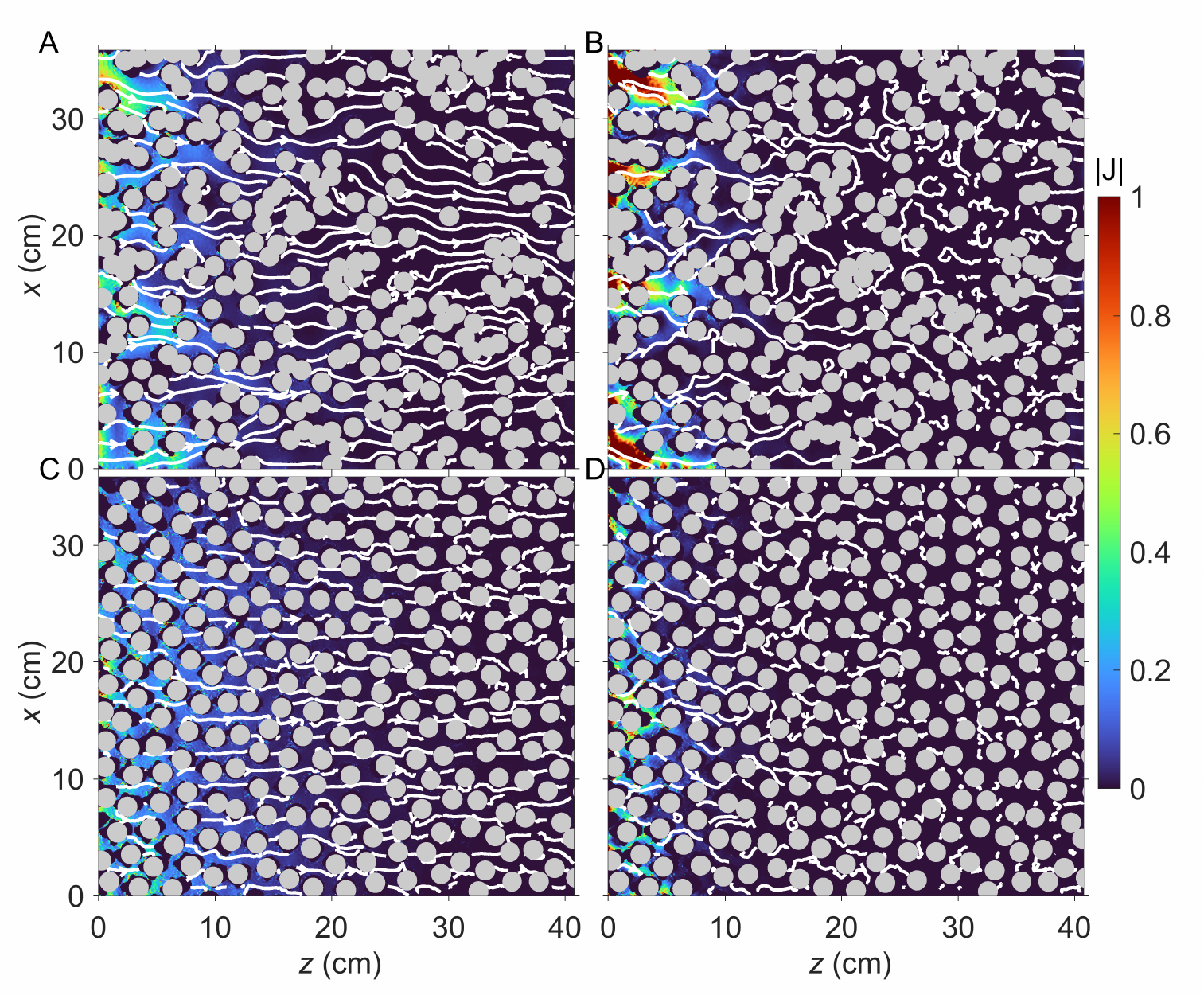}
\caption{Energy current and streamlines. (A)~Uncorrelated medium at the excitation frequency $f=\unit{5.0}{Hz}$, (B)~at $f=\unit{8.0}{Hz}$. (C)~SHU medium at the excitation frequency $f=\unit{5.0}{Hz}$, corresponding to the non-scattering regime, and (D) at $f=\unit{8.0}{Hz}$ in the strong scattering regime. Colors indicate $|\mathbf{J}|$ (arbitrary units).}
\label{fig:Poynting}
\end{figure}

Taking advantage of the ability to measure the wavefield amplitude and its gradient in the Fourier checkerboard demodulation process, we calculate the normalized energy current $\mathbf{J} = \Im(\psi^* \nabla \psi)$ that characterizes the local strength and direction of the energy flow~\cite{carminati_principles_2021}. Maps of $\mathbf{J}$ within the scattering medium are shown in Fig.~\ref{fig:Poynting}, together with isolines of equal flux, for both the uncorrelated and SHU media and for two excitation frequencies below and above the threshold value. In the uncorrelated medium at 5~Hz (Fig.~\ref{fig:Poynting}A), the flux lines display spatial variations and smooth bending. In contrast, the SHU medium at the same frequency (Fig.~\ref{fig:Poynting}C) exhibits much straighter and more uniformly oriented flux lines, another signature of the suppressed scattering. At the higher frequency 8~Hz, outside the non-scattering region, the current density becomes more tortuous for both the uncorrelated medium (Fig.~\ref{fig:Poynting}B) and the SHU medium (Fig.~\ref{fig:Poynting}D).

\section*{Discussion}

Our results reveal two distinct regimes for wave transport in SHU disordered media: Suppression of scattering for small wavenumbers $k \leq K/2$, with $K$ denoting the size of the stealthy region in the structure factor, and strong scattering for $k > K/2$. A feature of stealth hyperuniformity is that in the non-scattering regime $k \leq K/2$, an uncorrelated medium with the same density of scatterers remains strongly scattering. Suppression of scattering by SHU differs from a mere homogenization process and does not coincide with an asymptotic regime in which all structural length scales are made much smaller than the wavelength.

The different regimes are visually highlighted by direct imaging of the wavefield inside the structure (Fig.~\ref{fig:wavefield}). By measuring the extinction length and the viscous absorption length in the background medium, we define an effective scattering coefficient that exhibits a clear transition between the non-scattering and strong scattering regimes. The transition occurs exactly at the predicted threshold wavenumber $K/2 = \unit{1.12}{cm^{-1}}$ (Fig.~\ref{fig:l_e}). Below this threshold, the effective scattering strength is reduced by a factor greater than 1.5 in the SHU medium, with extinction dominated by absorption in this ``transparency'' regime~\cite{Bigourdan19,Sheremet20}. Above threshold, scattering emerges in the SHU medium, and the effective scattering strength takes similar values for both uncorrelated and SHU media. The distinct regimes below and above the threshold wavenumber are also clearly visible in the fluctuations of the effective scattering strength (Fig.~\ref{fig:l_e std}), where self-averaging is observed for the SHU medium below the threshold value. The differences in wave transport below and above the threshold wavenumber are also highlighted in the energy current flux lines (Fig.~\ref{fig:Poynting}), where directivity and coherence of the wavefront are preserved only in the SHU medium below the scattering threshold.

 Overall, our measurements provide the first in-medium, quantitative demonstration of the transport regimes predicted for SHU disorder in water waves. By directly accessing the wavefield over a broad spectral range, we show that SHU correlations suppress scattering and yield deterministic transport behavior at the level of individual realizations, even in a lossy background. The observed non-scattering window validates theoretical predictions, and establishes hyperuniformity as an experimentally accessible design principle for controlling wave propagation across all classes of waves.
 
\section*{Materials and Methods}

\subsection*{Water tank and wave excitation}
Experiments are carried out in a $150 \times 60\,\mathrm{cm}$ water tank filled with a $6.9$--$7\,\mathrm{cm}$ layer of tap water. Surface waves are generated by a $40\,\mathrm{cm}$-wide paddle driven by a loudspeaker. A sinusoidal signal from a function generator (Rigol DG4162), amplified by an audio amplifier (König AMP 4800), produces monochromatic waves in the frequency range $3.5$--$10\,\mathrm{Hz}$. The corresponding wavelengths span $2.3$ to $14\,\mathrm{cm}$, covering the expected transition between non-scattering and strong scattering for SHU media. The paddle width is large compared to the wavelength, allowing the generated waves to be approximated as planar within the observation window. A sloped absorbing beach at the opposite end of the tank suppresses unwanted reflections.

\subsection*{Generation of SHU point patterns}
The SHU point patterns were generated following the reciprocal-space optimization procedure described in Ref.~\cite{Monsarrat22}. Periodic boundary conditions were imposed, such that the system is fully characterized by a discrete set of {\bf q} vectors in reciprocal space. Consequently, hyperuniformity can be enforced by constraining only a finite number of {\bf q} vectors. Stealthiness is achieved by imposing $S(q) = 0 \ \mathrm{for} \ q\leq K$. Starting from a Poisson distribution of points, their positions ${\bf r}_j$ are optimized by minimizing the objective function
\begin{equation}
U = \sum_{q \leq K} \left | \sum_{j=1}^N \exp(-i{\bf q}\cdot {\bf r}_j) \right |^2 \, .
\end{equation}
This quantity corresponds to the total spectral weight contained within the constrained region of reciprocal space. Since each term in the sum is non-negative, the global minimum $U=0$ is reached only when all Fourier amplitudes vanish inside the constrained region. The optimization therefore progressively drives the system towards a SHU state. In practice the minimization is performed using a gradient-descent algorithm until convergence was achieved. For the structures used in this work, with $K = 2.24\,\mathrm{cm}^{-1}$, the optimization converged to a final objective value $U=6.37 \times 10^{-8}$, indicating that the stealthiness constraints were satisfied to high numerical accuracy. A discussion of finite-size effects and the difference between SHU and hard-disk patterns is included in the SI Appendix (Fig. S2).

\subsection*{Fabrication of the disordered media}
The scattering media are composed of cylindrical PMMA scatterers ($1\,\mathrm{cm}$ diameter, $8\,\mathrm{mm}$ height) inserted into a $45 \times 45\,\mathrm{cm}$ transparent PMMA plate of thickness $8\,\mathrm{mm}$. Circular holes are laser-cut to host the scatterers, which are non-resonant and subwavelength in the explored frequency range. Two kinds of spatial distributions of scatterers with identical density 
$\rho = 0.2\,\mathrm{cm}^{-2}$ ($N = 401$ scatterers) are fabricated. The uncorrelated media are generated via random sequential addition. Candidate positions are drawn from a uniform distribution and accepted only if they satisfy a non-overlap condition with previously placed scatterers. This produces a random arrangement with minimal short-range correlations. The positions of the scatterers for the SHU media are generated using the procedure summarized in the previous section. Each PMMA plate can be rotated by $0^\circ$, $90^\circ$, $180^\circ$ or $270^\circ$, providing four statistical realizations for ensemble averaging.

\subsection*{Wavefield acquisition, reconstruction and complex-field extraction}
A camera (Basler acA1300-200\,\textmu m, 16\,mm/F1.4 lens) positioned $2\,\mathrm{m}$ above the tank records the deformation of a checkerboard placed beneath the transparent bottom. Videos are recorded at $35$--$100\,\mathrm{fps}$, ensuring at least ten frames per wave period. Each acquisition contains $200$ frames ($\approx 20$ periods).

\subsection*{Masking of the scatterers}
Before reconstruction, scatterers are removed from the raw images by thresholding the gradient of a reference frame and dilating the resulting binary mask by approximately $5$ pixels. The mask is applied prior to the demodulation stage to eliminate near-field artefacts. For the visualisation of the field and the mask, a different method is used. For the SHU medium, centers of the scatterers are found by using a circular Hough transform method, and then placing large enough circular disks to mask the scatterers at these positions. For the uncorrelated medium, the positions are recorded manually.

\subsection*{Extracting phase and amplitude}
We apply a spatial high-pass filter to remove the spurious low-frequency components generated by the reconstruction method. A lock-in technique is then applied by computing the Fourier component at the excitation frequency, enabling us to extract the amplitude and phase of the wave at that frequency. This procedure is performed for the total field (scatterer present) and for the reference incident field (cylinder absent). The scattered field is obtained by adjusting the phase of the incident wave and subtracting it from the total field.
 
\begin{acknowledgments}
This work has received support under the program ``Investissements d’Avenir'' launched by the French Government. M. Filoche and A. Campaniello are supported by the project Localization of Waves of the Simons Foundation (Grant No. 1027116, M.F.). E. Fort is supported by the AXA Research Fund. We thank Romain Pierrat and Quentin Louis for insightful discussions.
\end{acknowledgments}

\bibliography{biblio}

\end{document}